\documentclass[11pt,letterpaper]{article}
\usepackage[top=0.85in,left=0.75in,footskip=0.75in]{geometry}
\usepackage[utf8]{inputenc}
\usepackage{cite}
\usepackage{nameref,hyperref}
\usepackage[right]{lineno}
\usepackage{microtype}
\DisableLigatures[f]{encoding = *, family = * }
\usepackage{setspace} 
\usepackage{changepage}

\usepackage[labelfont=bf,labelsep=period,singlelinecheck=off,font={footnotesize,sf}]{caption}
\makeatletter
\renewcommand{\@biblabel}[1]{\quad#1.}
\makeatother

\usepackage{lastpage,fancyhdr,graphicx}
\usepackage{float}
\usepackage{color}

\usepackage{sidecap}

\begin{document}
\begin{flushleft}
{\huge
\textbf\newline{Photoacoustics can image spreading depolarization \\deep in gyrencephalic brain
}
}
\newline
\\
Thomas Kirchner \textsuperscript{1,2,*}, 
Janek Gröhl \textsuperscript{1,3}, 
Mildred A. Herrera \textsuperscript{4}, 
Tim Adler \textsuperscript{1,5}, 
Adrián Hernández-Aguilera \textsuperscript{4}, 
Edgar Santos \textsuperscript{4}, 
Lena Maier-Hein \textsuperscript{1,3,\ddag}
\\
\bigskip
{\small
\bf{1} Division of Computer Assisted Medical Interventions, German Cancer Research Center, Heidelberg, Germany.
\\
\bf{2} Faculty of Physics and Astronomy, Heidelberg University, Heidelberg, Germany.
\\
\bf{3} Medical Faculty, Heidelberg University, Heidelberg, Germany.
\\
\bf{4} Department of Neurosurgery, Heidelberg University Hospital, Heidelberg, Germany.
\\
\bf{5} Faculty of Mathematics and Computer Science, Heidelberg University, Heidelberg, Germany.
\\
\bigskip
*\, t.kirchner@dkfz-heidelberg.de
\\
\ddag\, l.maier-hein@dkfz-heidelberg.de
}
\end{flushleft}

\section*{Abstract}
Spreading depolarization (SD) is a self-propagating wave of near-complete neuronal depolarization that is abundant in a wide range of neurological conditions, including stroke. SD was only recently documented in humans and is now considered a therapeutic target for brain injury, but the mechanisms related to SD in complex brains are not well understood. While there are numerous approaches to interventional imaging of SD on the exposed brain surface, measuring SD deep in brain is so far only possible with low spatiotemporal resolution and poor contrast.
Here, we show that photoacoustic imaging enables the study of SD and its hemodynamics deep in the gyrencephalic brain with high spatiotemporal resolution. As rapid neuronal depolarization causes tissue hypoxia, we achieve this by continuously estimating blood oxygenation with an intraoperative hybrid photoacoustic and ultrasonic (PAUS) imaging system. Due to its high resolution, promising imaging depth and high contrast, this novel approach to SD imaging can yield new insights into SD and thereby lead to advances in stroke, and brain injury research.

\section*{Main}
Spreading depolarization (SD) is a self-propagating wave of near-complete neuronal depolarization that occurs abundantly \cite{dreier_role_2011} in individuals with progressive neuronal injury after stroke \cite{dohmen_spreading_2008} and traumatic brain injury \cite{hartings_spreading_2011} as well as subarachnoid hemorrhage \cite{dreier_delayed_2006}, intracerebral hemorrhage \cite{helbok_clinical_2017}, and migraine with aura \cite{lauritzen_cerebral_1987, hadjikhani_mechanisms_2001}. Sixty years after the discovery of SD \cite{leao_spreading_1944}, many mechanisms related to SD have still not fully been understood while recent research increasingly finds SDs to be a therapeutic target in injured brain \cite{ayata_spreading_2015, chung_spreading_2016}.

In order to increase understanding of SD, the morphologies of their wave fronts have been a subject of intense study \cite{woitzik_propagation_2013, takano_cortical_2007, santos_heterogeneous_2017}. In these, the gyrencephalic brain has been found to be capable of irregular SD propagation patterns \cite{santos_heterogeneous_2017, santos_radial_2014} not found in lissencephalic brain. It remains to be studied if and how these patterns occur and evolve in depth. A better classification of the morphologies of these wave fronts could lead to a clear definition of a therapeutic target beyond mere occurrence of SD. The methods used to study SD can be classified in electrophysiological and optical approaches. The current clinical state of the art for monitoring SD is electrocorticography (ECoG) using subdural electrodes placed directly on the cortex \cite{dreier_recording_2017, hartings_spreading_2014} to record electrical activity. Because SDs propagate far from their point of origin, placing ECoG electrodes allows for remote monitoring of various brain injury. Characteristic patterns usually appear delayed for adjacent electrodes, with an SD registering as a large near direct current (DC) shift in the electrodes signal, followed by persistent depression of spontaneous cortical activity registering as higher frequency alternating current (AC) signal components \cite{dreier_recording_2017}. While ECoG is clinical practice for surface measurements,  implanting electrodes deep into the brain is the prime method of investigating SD beyond the brain surface. Doing so, SDs have been shown to occur in deep structures of the lissencephalic brain and in the brainstem, where they have been associated with sudden unexpected death in epilepsy \cite{aiba_spreading_2015}. How an SD, which originates on the cortex spreads to deep structures without direct gray matter connection is unclear as the use of electrical monitoring does not yield sufficient spatial information. 

While optical techniques are not in routine clinical use, a range of them are used to study SD. They can yield high spatiotemporal resolution information usually related to the hemodynamic response to SD. These techniques include two photon microscopy (TPM) \cite{chuquet_high-resolution_2007, takano_cortical_2007,murphy_two-photon_2008,risher_recurrent_2010}, laser speckle (LS) imaging \cite{jones_simultaneous_2008, dunn_dynamic_2001,woitzik_propagation_2013}, intrinsic optical signal (IOS) imaging \cite{ba_multiwavelength_2002, santos_cortical_2013,santos_heterogeneous_2017} and near infrared spectroscopy (NIRS) \cite{seule_hemodynamic_2015,winkler_impaired_2012,mccormick_noninvasive_1991}. TPM has an exceptional, single cell spatial resolution using the fluorescence of reduced nicotinamide adenine dinucleotide (NADH) as contrast. It can achieve a temporal resolution in a seconds range for a sub-millimeter imaging field \cite{takano_cortical_2007} and has a sub-millimeter penetration depth. TPM is therefore mostly used in small animal models. LS imaging or LS flowmetry images changes in cerebral blood flow in single vessels \cite{dunn_dynamic_2001}. It is complementary to the larger field of view IOS imaging \cite{jones_simultaneous_2008} which images reflectance changes of light in one \cite{santos_cortical_2013} or two \cite{ba_multiwavelength_2002} narrow bands. IOS has approximately one second temporal resolution and micron spatial resolution, while again being diffusion limited to a sub-millimeter penetration depth and no depth resolution. NIRS, in contrast to the other optical techniques, is no imaging technique but employs point measurement probes \cite{seule_hemodynamic_2015} or optrode strips \cite{winkler_impaired_2012} to monitor millimeter scale areas similar to electrodes. Like IOS, it indirectly measures reflectances correlated to relative concentration changes of the chromophores oxyhemoglobin (HbO) and deoxyhemoglobin (Hb), but its signal response is integrated over the area under the probe, leading to single spot measurements and no spatial resolution.

Functional magnetic resonance imaging (fMRI) with blood oxygen level dependent (BOLD) or diffusion weighted contrasts is the only  modality that has been used to image the hemodynamic response of SD deep in brain \cite{hadjikhani_mechanisms_2001}. Substantial drawbacks besides the complex imaging setup are the poor spatiotemporal resolution \cite{umesh_rudrapatna_measurement_2015,yao_photoacoustic_2014} and low contrast \cite{hadjikhani_mechanisms_2001} when compared to optical or electrical measurements.

Overall, it can be concluded that the imaging methods proposed to date either feature high spatiotemporal resolution (IOS, TPM, LS) or are capable to provide depth-resolved information on SD beyond the surface (fMRI, implanted electrodes), but cannot provide both. To address this bottleneck, we investigate photoacoustic (PA) imaging as a possible high-resolution imaging technique for measuring SD deep in the gyrencephalic brain. Near infrared (NIR) light can penetrate deep into tissue, is scattered and gets diffused, thereby losing spatial information after a fraction of a millimeter. Photoacoustics \cite{wang_photoacoustic_2012} is capable of imaging beyond this sub-millimeter optical diffusion limit through the PA effect \cite{bell_art_1880}; light is delivered as a nanosecond laser pulse and where it is absorbed, it causes sudden thermoelastic expansion which in turn gives rise to acoustic waves. The ultrasonic spectral component of these waves emitted by the PA effect scatter much less than NIR light in tissue and can be detected by ultrasound (US) probes. Reconstructing their origin yields PA images. A multispectral stack of such images can be processed to reconstruct images of estimated tissue oxygenation that feature  the spatiotemporal resolution and imaging depth of US combined with the optical contrast of NIRS. Multispectral photoacoustic imaging has shown to image blood oxygenation and perfusion in a variety of applications \cite{xia_photoacoustic_2014,li_photoacoustic_2009,taruttis_advances_2015,tzoumas_spectral_2017,tzoumas_eigenspectra_2016}. In the context of brain imaging, however, the application of PA has been restricted to lissencephalic brains \cite{yao_photoacoustic_2014} and its potential for monitoring SD remains to be investigated. 

\begin{SCfigure}[][hbt!]
  \includegraphics[]{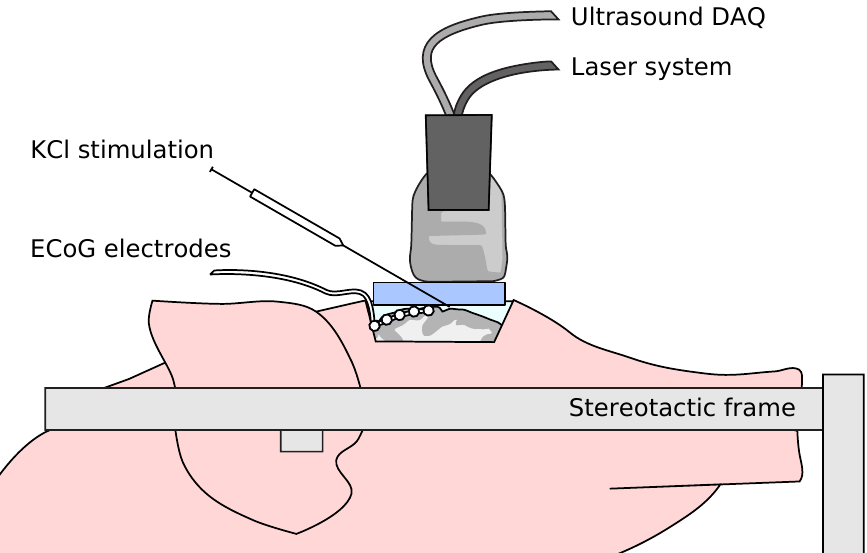}
  \caption{Setup for characterizing spreading depolarization (SD) deep inside the gyrencephalic brain with a hybrid photoacoustic ultrasonic (PAUS) imaging system. The PAUS probe is  placed on a gel pad to allow for PAUS-guided potassium chloride (KCl) stimulation in the imaging plane. Electrocorticography (ECoG) recordings serve as a clinical reference.}
  \label{fig1} 
\end{SCfigure}

Rapid neuronal depolarization and repolarization causes tissue hypoxia \cite{takano_cortical_2007}. Therefore, our work is based on the assumption that the imaging of hemodynamic changes with photoacoustics enables the monitoring of SD deep inside the tissue. We hypothesize that multispectral PA imaging is able to image SD induced hemodynamic changes in the entirety of the cortical gray matter of a gyrencephalic brain. For the purposes of this study, we measure an estimation of blood oxygenation (sO$_2$) and total hemoglobin (THb). Our imaging concept, which is illustrated in Fig.\,\ref{fig1}, relies on a hybrid photoacoustic ultrasonic (PAUS) imaging system which combines (1) an US research system featuring a linear US transducer with a  center frequency of 7.5\,MHz and broad acoustic response \cite{kirchner_freehand_2016} with (2) a near infrared (NIR) fast tuning optical parametric oscillator (OPO) laser \cite{kim_programmable_2016} (see Methods). The system operates in an interleaved PAUS imaging mode, acquiring multispectral PA sequences with corresponding US images for each PA image. This concurrent US imaging is used as anatomical reference for the physician (e.g. as guidance for the stimulation) and for motion compensation of the PA data (see Methods). Each multispectral PA image stack is converted into an image of estimated sO$_2$ and by spectral unmixing. 

\section*{Experiments \& Results}
Two experiments were performed with our PAUS system to investigate whether the monitoring of tissue oxygenation with PA enables the detection and monitoring of SDs in the entire depth of the cortical gray matter of a gyrencephalic brain. In both experiments, brain activity was monitored with ECoG using a standard subdural electrode strip (Fig.\,\ref{fig1}).

\begin{SCfigure}[][ht!]
    \includegraphics[]{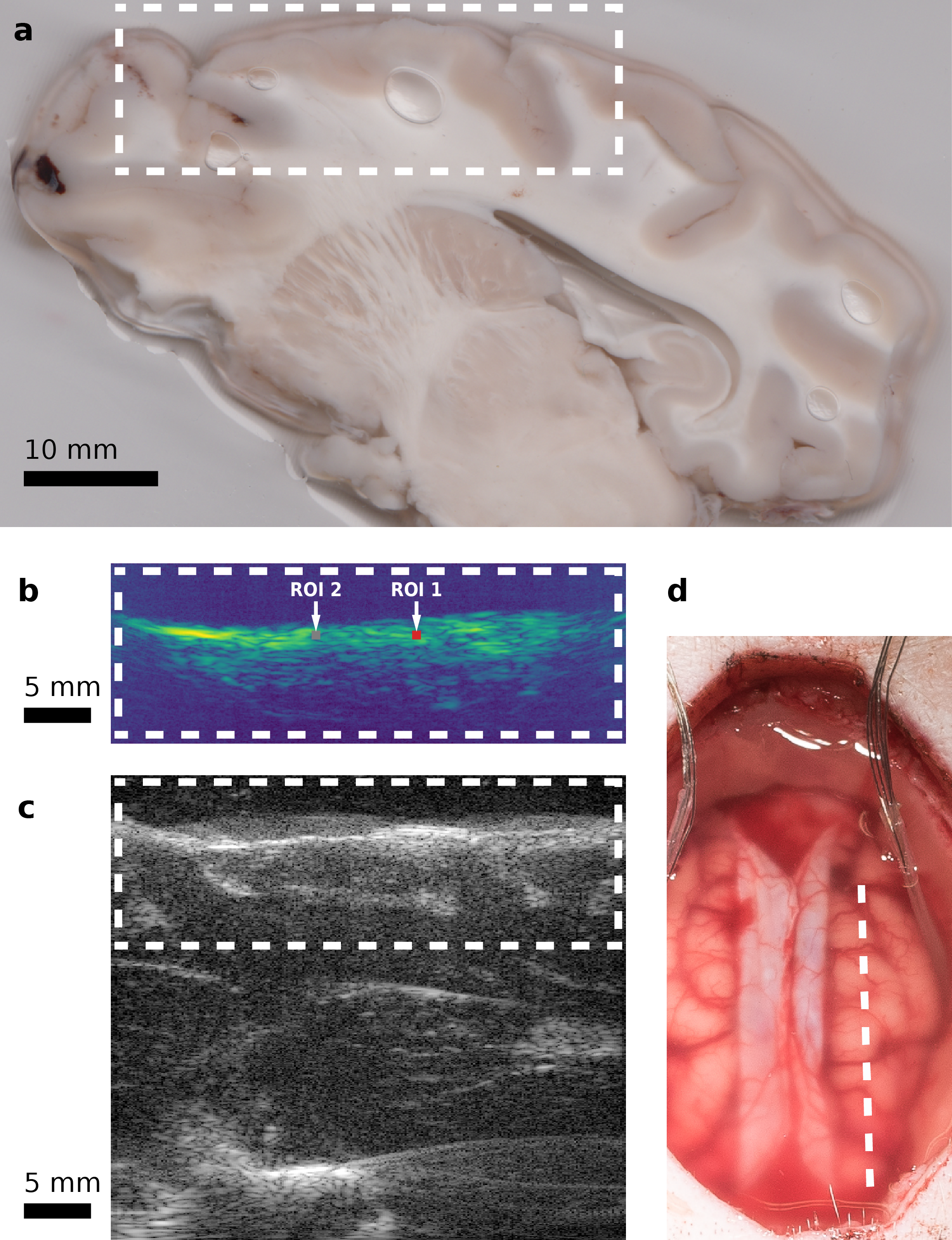}
    \caption{Hybrid photoacoustic ultrasonic  (PAUS) imaging of a porcine brain. The dashed white line and boxes show corresponding sections of the swine cortex. \textbf{(a)} Photograph of a sagittal surgical slice segmented from the extracted brain, 1\,cm from the midline. The shown segment is manually registered to \textbf{(b)} a representative photoacoustic (PA) image with two regions of interest (ROI), and \textbf{(c)} the corresponding ultrasound (US) B-Mode image. \textbf{(d)} Photograph of the exposed cortex after the craniotomy and dura mater retraction with the dashed line marking the PAUS imaging plane. The electrocorticography (ECoG) electrodes are positioned in the lateral margins.}
    \label{fig2} 
\end{SCfigure}

The aim of the \emph{initial wave experiment} was to investigate if the hemodynamic response of the brain to an induced SD can be imaged with multispectral PA. We performed the experiment in an uninjured brain. To analyze tissue hemodynamics before, during and after the occurence of SD, we took continuous PAUS measurements starting 24\,min before the the first potassium chloride (KCl) stimulation (see Fig.\,\ref{fig1}) and ending one hour after the stimulation. After the experiment we cut sagittal surgical slices from the extracted brain to relate the  acquired PA and US images to the brain morphology as seen on the exposed tissue. As shown in Fig.\,\ref{fig2} we were able to image PA signal up to a depth of approximately 1\,cm, which allowed us to image the entire cortical gray matter in the field of view of the imaging plane.

\begin{SCfigure}[][th!]
\includegraphics[]{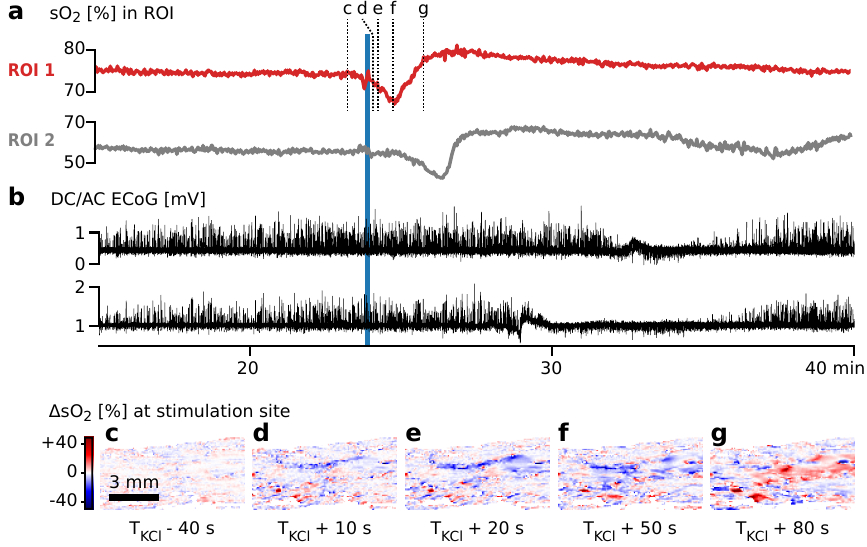}
\caption{Results of \emph{initial wave experiment} showing spreading depolarization (SD) starting from an equilibrium state. \textbf{(a)} Estimated blood oxygenation (sO$_2$) of two regions of interest (ROI) in the left hemisphere (see Fig.\,\ref{fig2}\,b).
\textbf{(b)} Simultaneous electrocorticography (ECoG) monitoring. Data from two adjacent  electrodes on the left hemisphere is shown -- the other three channels on the left hemisphere and the five channels on the right hemisphere showed no change. The electrodes were placed on the lateral margins of the brain as to not interfere with hybrid photoacoustic ultrasonic (PAUS) imaging. \textbf{(c)}--\textbf{(g)} Absolute change in estimated sO$_2$ ($\Delta$ sO$_2$) in a region near the stimulation site (c) before potassium chloride (KCl) stimulation and (d)--(g) 10-80\,s after stimulation. In (d)--(f) spreading, intensifying hypoxia is measured followed by (g) an overcompensation in sO$_2$.}
\label{fig3}
\end{SCfigure}

By estimating sO$_2$ in each pixel of our reconstructed multispectral images we observed a single wave of hypoxia spreading from the point of KCl stimulation through the tissue at a speed of approximately 5\,mm/min. The estimated sO$_2$ for two sub-surface regions of interest (ROI) is plotted in Fig.\,\ref{fig3}\,a to illustrate this wave. sO$_2$ in a wide field of view during the same time frame is shown in \nameref{Supplemental Video 1} played at a factor 100 timelapse. The wave of hypoxia coincides with the ECoG measurements on two electrodes in the proximity whose signals are plotted in Fig.\,\ref{fig3}\,b; they clearly show a single SD wave moving through the cortex, while the other electrodes on both hemispheres showed no change in activity. Fig.\,\ref{fig3}\,c--g shows the change in estimated sO$_2$ in the region around the stimulation as hypoxia propagating through the tissue followed by an increase in sO$_2$ over the baseline.

\begin{figure}[b!]
\includegraphics[]{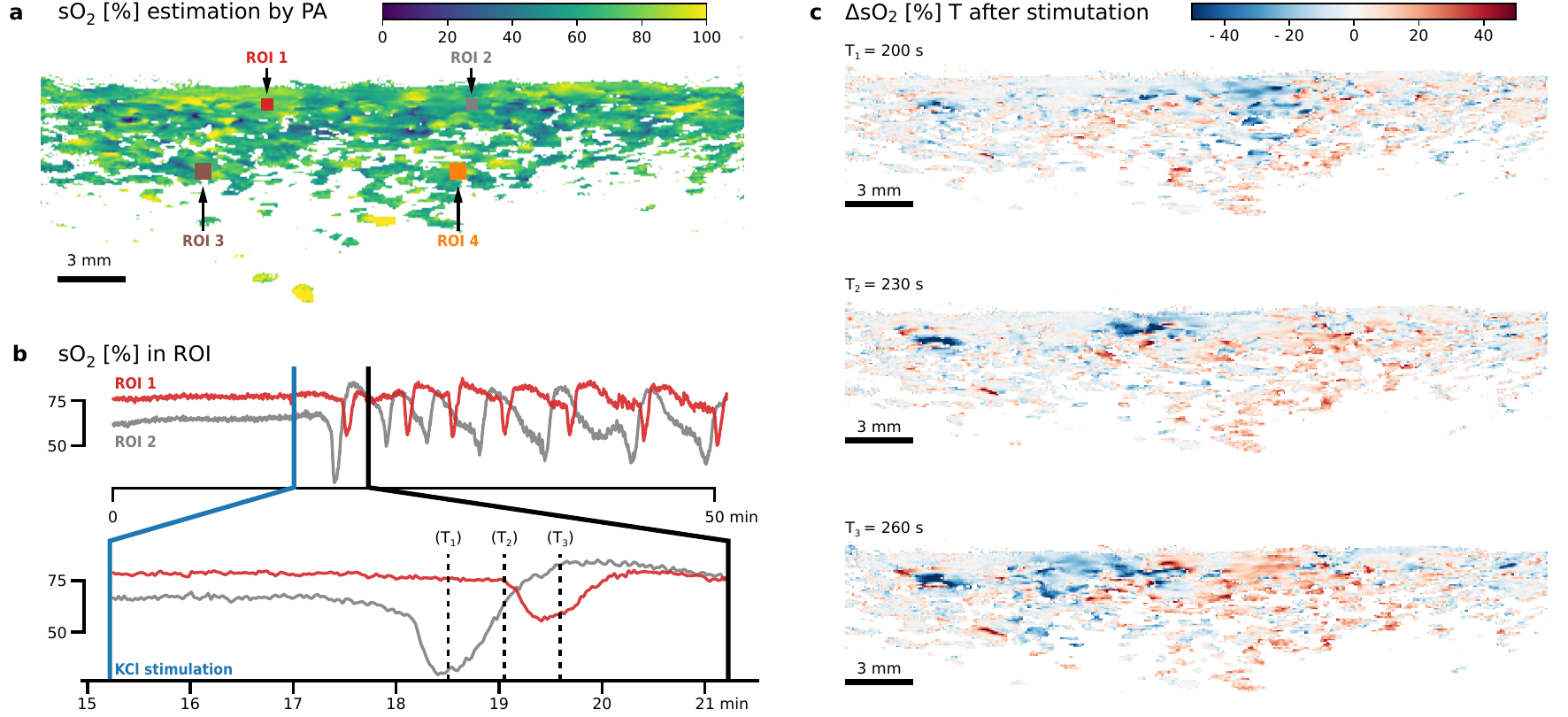}
\caption{Multispectral photoacoustic (PA) imaging of blood oxygenation (sO$_2$) as part of the \emph{cluster experiment}. After a 15\,min baseline scan, spreading depolarization (SD) was induced by potassium chloride (KCl) stimulation in the left hemisphere of a porcine brain. The sagittal plane was continuously imaged for 51\,min. \textbf{(a)} PA sO$_2$ estimation before stimulation with marked regions of interest (ROI). Refer to \nameref{Supplemental Video 2} -- a time lapse video of the change of sO$_2$ -- for a complete view. The playback speed is 90 times the recording speed. \textbf{(b)} Time evolution of estimated sO$_2$ in the two ROI (top: whole recording period; bottom: first wave). \textbf{(c)} The change in blood oxygenation ($\Delta$sO$_2$) relative to before KCl stimulation is shown for three example time steps 30 seconds apart (T$_1$, T$_2$ and T$_3$), corresponding to the dashed lines in (b).}
\label{fig4}
\end{figure}

The purpose of the \emph{cluster experiment} was to investigate the hemodynamic changes during SD clusters with PAUS. To this end, we repeatedly stimulated the brain with KCl until we observed the occurrence of clustered SDs both electrically with ECoG and optically with IOS as an additional state of the art reference for surface measurement of SD. Once the SDs had subsided, we started PAUS measurements with a new KCl stimulation  after a baseline recording period of 15\,min (details see Methods).

\begin{SCfigure}[][ht!]
\includegraphics[]{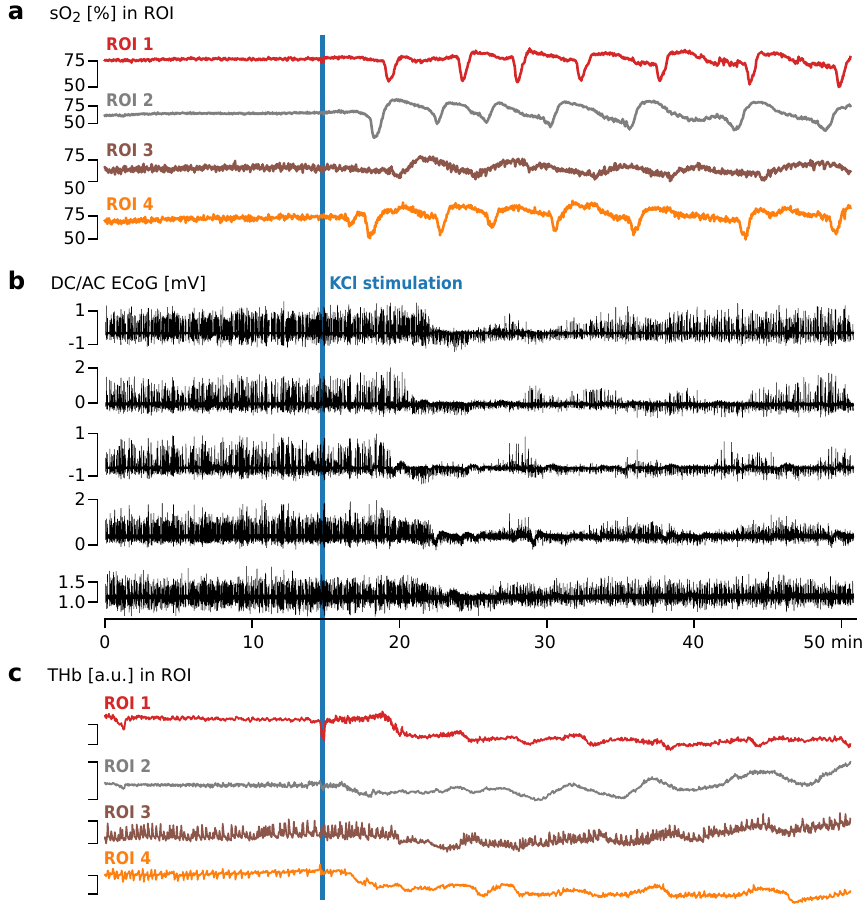}
\caption{Monitoring of hemodynamic changes in four regions of interest (ROI) (see Fig.\,\ref{fig4}\,a) as part of the \emph{cluster experiment}. Spreading Depolarization (SD) was induced 15\,min after start of recording by potassium chloride (KCl) stimulation on the left hemisphere of a porcine brain. (a) Blood oxygenation (sO$_2$)  in four ROIs in the left hemisphere.
(b) Simultaneous electrocorticography (ECoG) monitoring with five electrodes placed on the left hemisphere of the porcine brain. The occurrence of clustered SD is clearly visible as sudden direct current (DC) shifts spreading to neighboring channels, coinciding with spreading depression of the high frequency components.
(c) Monitoring of total hemoglobin (THb) in four ROI. In ROI\,3 and ROI\,4 low frequency vascular fluctuations (LF-VF) can be observed which appear to be periodically depressed by SD.}
\label{fig5}
\end{SCfigure}

As shown in Fig.\,\ref{fig4}\,a\,+\,b as well as \nameref{Supplemental Video 2}, there again was no change in the sO$_2$ estimation in the imaged sagittal plane in the left hemisphere during the baseline period. After KCl stimulation, we observed repetitive waves of hypoxia propagating through the imaging plane, followed by an overcompensation in sO$_2$ propagating through the cortex to up to a depth of approximately 2 to 5\,mm below the brain surface. Fig. \ref{fig4}c illustrates one such wave propagating from left to right during one minute as a change in sO$_2$ estimation. The speed of the waves was measured as 3--9\,mm/min between ROI\,1 and 2. ECoG measurements on the left hemisphere shown in Fig.\,\ref{fig5}\,b indicate a SD cluster with the same frequency and speed of the sO$_2$ changes (Fig.\,\ref{fig5}\,a). As was the case in the \emph{initial wave experiment} no change in ECoG activity in the right hemisphere was observed.

In addition to the sO$_2$ estimation from spectral unmixing we estimated the total hemoglobin (THb) for the \emph{cluster experiment}; this is visualized in \nameref{Supplemental Video 3}. The changes of THb in the ROI are shown in Fig.\,\ref{fig5}\,c, where ROIs\,3 and 4 seem to exhibit low frequency vascular fluctuations (LF-VF) \cite{obrig_spontaneous_2000} which appear to be depressed after SD \cite{dreier_cortical_2009,woitzik_propagation_2013}.

\section*{Discussion}
We investigated the imaging of SDs based on the concept of PA imaging. Our approach involves simultaneous US and multispectral PA imaging for time-resolved reconstruction of tissue oxygenation in sagittal image slices. Two \emph{in vivo} porcine experiments with our PAUS system provide the following evidence suggesting that our concept allows for the detection and monitoring of SD.

(1) \emph{Hypoxia consistent with ECoG:} By estimation of sO$_2$, we observed pronounced drops in estimated sO$_2$ after KCl stimulation (cf.\,Fig.\,\ref{fig3}\,a\,+\,\ref{fig4}\,b). This local hypoxia lasted for around 30 seconds and was followed by an overcompensation or return to baseline sO$_2$. These changes were consistent with the occurrence of SD in ECoG. The indicators we used to identify SD in ECoG were based on consensus \cite{dreier_recording_2017}: A characteristic abrupt DC shift followed by a longer lasting positivity, and a reduction in amplitudes of spontaneous AC activity. Both of which needed to spread with a speed of 1.5--9.5\,mm/min between electrodes and not cross hemispheres.

(2) \emph{Transient increase in blood volume:} By estimation of THb, we also observed a so-called \emph{normal hemodynamic response} -- a pronounced transient increase in blood volume (hyperemia) which was followed by a mild long-lasting oligemia (Fig.\,\ref{fig5}\,c). Note that the results from THb estimation are less conclusive compared to the sO$_2$ based measurements as THb estimations are more susceptible to absolute changes in light fluence over all measured wavelengths. This is caused by the change in illumination geometry due to swelling which causes a slow shift in the absolute THb signal as shown in Fig.\,\ref{fig4}\,c. Lower fluence, generally in higher depths also cause the signal to noise ratio to deteriorate, which can be observed when comparing ROI\,1 with ROI\,3 in Fig.\,\ref{fig5}\,a.

(3) \emph{Speed of wave propagation:} Both the changes in sO$_2$ and THb propagated through the gray matter at speeds of 3--9\,mm/min. This is consistent with speed of SD reported in the literature as 1.7--9.2\,mm/min \cite{woitzik_propagation_2013} or 1.5--9.5\,mm/min \cite{dreier_recording_2017} in gyrencephalic brain (3--9\,mm/min in porcine brain \cite{santos_heterogeneous_2017}).

(4) \emph{Low-frequency vascular fluctuations:} We also observed changes in low‐frequency vascular fluctuations (LF‐VF) \cite{dreier_cortical_2009} (Fig.\,\ref{fig5}\,c). The observed LF‐VF "display[ed] a spreading suppression in a similar fashion to that of SDs" in ECoG (see \cite{winkler_impaired_2012}). Note that LF-VF were only visible in the vicinity of larger vessels, which was the rationale for placing ROIs\,3 and 4 in such regions. 

We conclude from these observations that our measurements clearly support our initial hypothesis and suggest that PAUS is able to image SD as a change in sO$_2$. In contrast to all other methods proposed to monitor SD to date, our approach has the unique advantage that it features both high resolution and high imaging depth. While it is not suitable for imaging the entire gyrencephalic brain, penetration depth is sufficient to image the entire thickness of the cortical grey matter. Furthermore, the simultaneous PA and US imaging proved to be useful for anatomical orientation during the intervention (i.e. for needle guidance for KCl stimulation). Given these advantages, we see a potential use of PA imaging for SD characterization i.e. during pharmacological trials on the gyrencephalic brain. As the thickness of the human cerebral cortex is comparable, usually averaging 2.5\,mm and not exceeding 5\,mm \cite{fischl_measuring_2000}, PA imaging would be ideally suited for the study of SD in patients, as well. While, PA imaging cannot currently penetrate through an intact human skull\cite{yao_photoacoustic_2014}, PA imaging could for example be used postoperative to study SD in stroke patients \cite{lauritzen_clinical_2011} with a hemicraniectomy\cite{juttler_hemicraniectomy_2014}.

Our pilot study strongly suggests that photoacoustics could become a valuable tool for detection, imaging, and monitoring SD. Due to its high spatiotemporal resolution this approach can be used to more precisely study where (i.e. which neuron layer) SDs originate and how they propagate, thus adding to our understanding of the nature of SD and its contribution to brain injuries and disease progression.

\section*{Methods}
\subsection*{PAUS imaging system}
The custom built hybrid photoacoustic ultrasonic (PAUS) imaging system is based on a 128 channel ultrasound data acquisition system (DiPhAs, Fraunhofer IBMT, St. Ingbert, Germany) with a 128-element linear US transducer operating on a center frequency of 7.5\,MHz and broad acoustic response (L7-Xtech, Vermon, Tours, France). Due to its low level application programming interface (API) access, the system allows for raw data access and an interleaved PAUS imaging mode. This interleaved mode acquires US data from several shots after each PA data acquisition. The data acquisition (DAQ) module is combined with a fast tuning optical parametric oscillator (OPO) laser cart (Phocus Mobile, Opotek, Carlsbad, USA) which yields 690\,nm\,--\,950\,nm, 5\,ns long laser pulses with a pulse repetition rate of 20\,Hz and a per laser pulse power of up to 50\,mJ. The wavelength of each laser pulse can be tuned in between shots, allowing for real time multispectral acquisition sequences. Laser fiber bundles ending in two line arrays are attached to the transducer by a 3D-printed frame including acrylic windows for the laser output. For each experiment the entire probe was wrapped in a sterile ultrasound probe cover and gold leaf was placed between the US transducer and the probe cover to reduce artifacts created by light absorption in the US transducer. For live imaging and recording all APIs to the system were integrated in the Medical Imaging Interaction Toolkit (MITK) software framework and the MITK workbench application was used throughout the intervention to control the PAUS system, configure the image acquisition, and show live PA and US imaging streams. During our experiments we visualized both streams with 15--20\,fps using delay and sum (DAS) beamforming for an imaging depth of 4\,cm with 256 reconstructed lines. 
For the \emph{initial wave experiment} we imaged the wavelength sequencence (735\,nm, 756\,nm, 850\,nm, 900\,nm) selected to distinguish Hb and HbO \cite{luke_optical_2013}. Because we added an estimation of THb in the \emph{cluster experiment} we instead imaged the isosbestic point of Hb and HbO at 798\,nm for further reference, leading to the wavelength sequencence (760\,nm, 798\,nm, 858\,nm).

\subsection*{Image reconstruction}
The raw radiofrequency (rf) PA data acquired during the experiments was matched with the laser pulse energies recorded by a pyroelectric sensor (Ophir PE25-C, Ophir Optronics, North Andover, USA) built in the laser system (Phocus Mobile, Opotek, Carlsbad, USA) and matched with the wavelengths of the laser pulses measured by a spectrometer (HR2000+, Ocean Optics, Dunedin, USA). The wavelengths of the pulses were measured independently of the imaging system to account for calibration errors. The rf PA slice was then corrected for the corresponding pulse energies. The recorded PAUS data was already beamformed live during the experiment to reduce the system load writing to disk. A single US slice was recorded after each PA slice. The US image was a compounded image averaged from US data acquired at five angles, equidistant from $+10$\,deg to $-10$\,deg and beamformed to 256 lines using a delay and sum (DAS) algorithm with boxcar apodization.
To convert the acquired rf PA slices into meaningful images suitable for multispectral analysis, the slices were beamformed with a reference DAS implementation \cite{kirchner_signed_2018} using Hanning apodization to 512 lines. B-Mode images with isotropic pixel spacing of 0.075\,mm were formed with a Hilbert transform based envelope detection filter. US B-Mode images were formed in the same way, only adding a subsequent logarithmic compression.

\subsection*{Motion compensation}
The PA images obtained after beamforming are corrected for inter-frame motion introduced by breathing, pulse or swelling (1) to enable a more stable spectral unmixing and (2) to assure that a given pixel location corresponds to the same physical  regions of interest (ROI). To correct for the inter-frame motion an optical flow based method is employed. The optical flow of each US image relative to the first US image in the entire recording is estimated using an algorithm proposed by Farnebäck \cite{farneback_two-frame_2003}. The flow estimated from the US B-mode image is then used to warp the corresponding PA B-mode image.

\subsection*{Experimental Data Analysis}
Because of the slow propagation of SD wave fronts we averaged over ten motion corrected frames of the same wavelength and still have the 1\,s temporal resolution of IOS. Spectral unmixing of those image sequences was then performed using a non-negative constrained linear least squares solver (\verb!scipy.optimize.nnls!). In all figures and supplemental material plots and videos one PA datapoint is averaged over ten frames and then averaged over the ROIs. Speeds of SD wavefronts were obtained by measuring the time between the local minima of sO$_2$ ROI\,1 and ROI\,2. The positions of ROIs 1 and 2 in both experiments were chosen at 1\,mm depth and 7.5\,mm apart, in the center of the reconstructed image stream. ROIs\,3 and 4 were chosen deeper and close to larger vessels to investigate the LF-VF effect which, as discussed, can only be observed there. ROIs are otherwise representative of the entire data set as can be seen in the supplemental videos.

\subsection*{Animals}
Protocols for all experiments were approved by the institutional animal care and use committee in Karlsruhe, Baden-Wuerttemberg, Germany (Protocol No. 35-9185.81/G-174/16). Female German Landrace swines of 31 and 33\,kg were premedicated with Midazolam (Dormicum 0.5--0.7\,mg/kg) and Azaperone (Stresnil 4\,mg/kg) intramuscularly.  After premedication, two venous lines were placed in the ear veins, and propofol (Disoprivan 5--7\,mg/kg) was administered intravenously to facilitate the intubation. The animals were then intubated and mechanically ventilated and the pressure controlled ventilation was adapted to a respiration rate of 12--20/min, a flow of 2.5\,l O$_2$/min, 2.0\,l air/min, FiO2 35\,\% and volume 7--10\,ml per kg. The maintenance of anesthesia required inhalational anesthesia with isoflurane (Isosthesia 0.6–-1.0\,\%) and intravenous midazolam at a continuous dose of 0.5--0.7\,mg/kg/h via perfusion and maintained throughout the entire experiment.  If a wakening reaction occurred, a bolus of propofol (Disoprivan 5--7\,mg/kg) was administered.  Temperature was monitored with a rectal probe. A 4-Fr catheter was placed in the right femoral artery for permanent monitoring of the mean arterial blood pressure (Raumedic AG, Helmbrechts, Germany). Capillary oxygen saturation (SpO$_2$) was monitored from one ear. Arterial blood gases were obtained in the animal used for the \emph{initial wave experiment}. Ringer's solution was given intravenously over 8–-12\,h, to compensate for intraoperative bleeding, urinary output and insensible losses.
The two animals used in this study were used primarily for this project. After finishing the protocol the animal used for the \emph{cluster experiment} was used for other unrelated studies.

\subsection*{Surgery}
Animals were fixed in a stereotactic frame (Standard Stereotaxic Instruments, RWD Life Science, Shenzhen, China) and an extensive craniotomy with excision of the dura mater was performed, to view the subarachnoidal space bilaterally. Initially, the brain surface was immersed for 30 to 40\,min in a standard lactated Ringer's solution with an elevated K+ concentration (7\,mmol/l), as preconditioning for SD induction, as proposed by Bowyer et al. \cite{bowyer_analysis_1999} for the KCl model of SD. EcoG was performed with two strips of 5 electrodes each (Ad-tech, Racine, Wisconsin, USA) that were placed at the lateral margins of the craniotomy below the dura mater and above the parietal cortex. A camera for IOS imaging and its corresponding light sources were mounted above the stereotactic frame. After preconditioning a 5--10\,mm deep paraffin pool was filled over the exposed cortex, to reduce the diffusion of the KCl stimulation. When necessary, paraffin was withdrawn and new paraffin was added. The preparation time was 4--5\,h before the KCl stimulations started.
A gel pad (Aquaflex Ultrasound Gel Pad, Parker Laboratories, Fairfield, USA) was cut in shape of the exposed brain surface and placed in the paraffin pool. The custom designed PAUS probe (see Methods, PAUS Imaging system) was placed on top the gel pad and fixed relative to the frame. For the \emph{initial wave experiment} the gel pad and PAUS imaging system was placed before the initial stimulation. For the \emph{cluster experiment} the gel pad and system was placed after the initial KCl stimulations and the accompanying IOS imaging was performed. With the help of live US imaging, it was positioned to image a sagittal plane of the left hemisphere approximately 1\,cm from the midline. For the \emph{cluster experiment} we waited until any residual SD from prior stimulation subsided in the ECoG monitoring. Only then did we start recording PAUS data in a sagittal plane for 15\,min as a baseline. After a sufficient baseline recording, spreading depolarization was triggered using 2--5\,$\mu$l of 1\,mol/l KCl solution with a Hamilton syringe. The stimulus needle was guided using the live PAUS image streams visualized in MITK.

\subsection*{Monitoring}
All relevant physiological parameters, such as mean arterial pressure, rectal temperature, heart rate, and arterial oxygen saturation, were continuously monitored. A mean systolic arterial pressure of 60 to 80\,mmHg, a temperature between 36 and 37\,$^\circ$C, SaO$_2 > 90$\,\%, pCO$_2$ between 35 and 45\,mmHg, pO$_2 > 80$\,mmHg were maintained.

\subsection*{Electrocorticography}
Electrocorticography (ECoG) recording with the subdural electrodes was perfomed in 10 active channels, using the Powerlab 16/SP analogue/digital converter coupled with the LabChart-7 software (ADInstruments, New South Wales, Australia) at a sampling frequency of 400\,Hz.
For visualization, in all figures and supplemental material, ECoG data was post-processed in Python using a 45\,Hz Butterworth low pass filter to filter alternating current (AC) noise.

\subsection*{Intrinsic optical signal imaging}
intrinsic optical signal (IOS) imaging is a functional neuroimaging technique that measures cortical reflectance changes \cite{santos_cortical_2013}. We imaged one band at a wavelength of 564\,nm (14\,nm FWHM) with a charge-coupled device (CCD) camera (Smartec GC1621M, MaxxVision GmbH, Stuttgart, Germany) which was mounted 25\,cm above the exposed cortex. Images were acquired with static illumination and 2\,s CCD integration time. Changes in tissue reflectance were registered using a method described in \cite{santos_cortical_2013}. IOS was only used as an additional reference to the ECoG in the animal corresponding to the \emph{cluster experiment} to ensure that preconditioning was sufficient and SDs were easily triggered.



\section*{Acknowledgements}
The authors would like to acknowledge support from the European Union through the ERC starting grant COMBIOSCOPY under the New Horizon Framework Programme under grant agreement ERC-2015-StG-37960. The authors declare no conflict of interest.

\section*{Author Contributions}
T.K. conceived the study, implemented the system, designed the experiments, performed the experiments, analyzed the data, drafted the initial manuscript; J.G. designed the experiments, performed the experiments, analyzed the data, edited the entire manuscript; M.A.H. helped plan the experiments, performed the experiments, edited the entire manuscript; T.A. performed the experiments, helped analyze the data, edited the entire manuscript; A.H.-A. performed the experiments, edited the entire manuscript; E.S. conceived the study, designed the experiments, performed the experiments, supervised the neurosurgical aspects of the work, edited the entire manuscript; L.M.-H. conceived the study, designed the experiments, supervised the biomedical informatics and engineering aspects of the work, edited the entire manuscript.

\section*{Supplemental Material}
\setcounter{figure}{0}

\subsection*{Supplemental Video 1}
\label{Supplemental Video 1}
\emph{Initial wave experiment.} Blood oxygenation (sO$_2$): \href{https://goo.gl/wjnvav}{link to online video}

\subsection*{Supplemental Video 2}
\label{Supplemental Video 2}
\emph{Cluster experiment.} Blood oxygenation (sO$_2$): \href{https://goo.gl/CsEPvA}{link to online video}

\subsection*{Supplemental Video 3}
\label{Supplemental Video 3}
\emph{Cluster experiment.} Total Hemoglobin (THb) and low frequency vascular fluctuations (LF-VF): \href{https://goo.gl/RaUcbt}{link to online video}

\end{document}